\documentclass[letterpaper, conference]{IEEEtran}

\usepackage{graphicx}%
\usepackage{multirow}%
\usepackage{amsmath,amssymb,amsfonts}%
\usepackage{amsthm}%
\usepackage{stmaryrd}
\usepackage{xcolor}%
\usepackage{algorithm}%
\usepackage{algorithmicx}%
\usepackage{algpseudocode}%
\usepackage{listings}%

\usepackage{bm}

\usepackage{caption} 
\newcommand{\etal}[1]{#1~{\it et al.}}
\newcommand{\code}[1]{{\small\tt#1}}

\newcommand{\compose}{\bm{\circ}}

\newcommand{\Provided}{\textrm{provided}}

\newif\ifCorSep
\newcommand*{\corI}[1]{%
  \left[\relax
  \CorSepfalse
  \corIScan#1\relax\relax\relax
  \right]
}
\newcommand*{\corIScan}[1]{%
  \ifx\relax#1\empty
  \else
    \ifCorSep
     ;\relax
    \else
      \CorSeptrue
    \fi
    #1\relax
    \expandafter\corIScan
  \fi
}

\newcommand*{\corD}[1]{%
  \llbracket\relax
  \CorSepfalse
  \corIScan#1\relax\relax\relax
  \rrbracket
}

\newcommand*{\Yield}[1]{{!}#1}
\newcommand*{\Receive}[1]{{?}#1}
\newcommand{\Zero}{\mathbf{0}}

\newcommand*{\Start}{\operatorname{Start}}

\newcommand*{\Inline}{\operatorname{Inline}}

\newcommand{\Match}{\mathbf{match}}
\newcommand{\First}{\mathbf{first}}
\newcommand{\NotFound}{\mathbf{none}}
\newcommand{\Hd}{{\mathbf{h}}}
\newcommand{\Tl}{{\mathbf{t}}}

\newcommand{\Seq}[1]{{\left\langle#1\right\rangle}}
\newcommand{\Tuple}[1]{{\left ( #1 \right )}}

\newcommand{\tlit}[1]{\textrm{#1}}

\newcommand{\cpp}{\mbox{C{\tt++}}}
\newcommand{\Golang}{Golang}
\newcommand{\TestsCount}{21}

\newcounter{composing}
\newenvironment{composingEquation}{\refstepcounter{composing}\equation}{\tag{CR\thecomposing}\endequation}

\begin{document}

\newif\ifblind
\blindfalse

\title{A Flow Extension to Coroutine Types for \\Deadlock Detection in Go
\ifblind\else
\thanks{This study is supported by the Science and Technology Development Fund, Macau SAR, with File No.\ 0015/2023/RIA1.}\fi}

\ifblind
\author{\IEEEauthorblockN{Double Blind A\qquad Double Blind B\qquad Double Blind C}
\IEEEauthorblockA{
\textit{Double Blind Institute}\\
\textit{Double Blind University}, Earth \\
\{007, 008, 009\}@university.edu}}
\else
\author{\IEEEauthorblockN{Qiqi Jason Gu\qquad Lixue Liu\qquad Wei Ke}
\IEEEauthorblockA{
\textit{Macao Polytechnic University}\\
Macao SAR, China\\
qiqi.gu@cvut.cz, lixue.liu@mpu.edu.mo, wke@mpu.edu.mo}}
\fi

\maketitle

\begin{abstract}
Coroutines, as an abstract programming construct, are a generalization of functions that can suspend execution part-way for later resumption.
Coroutine Types are behavioral types to model interactions of coroutines with a single receiving operation followed by a single yielding operation.
Coroutine Types have been applied to model-driven engineering, smart contracts, and test case generation.
We contribute a Flow extension to Coroutine Types, so that coroutines with more than one receiving and yielding operation can be modeled.
We accordingly revise the reduction rules of Coroutine Types.
To show the usefulness of the Flow extension, we contribute a type system that maps expressions of the Go programming language to Coroutine Types.
If the reduction result is $\Zero$, the two channel operations are paired properly and the program has no deadlocks.

We choose Go because it is a popular programming language for distributed systems,
but a frequent kind of bugs in Go is deadlocks due to the wrong use of concurrency features.
We concentrate on the most commonly used semantics in Go: unbuffered channels with the keywords \code{go} and \code{defer}.
Our Flow extension and the type system recognize 17 patterns of channels and goroutine interactions,
including mismatched receivers and senders, nested goroutines, etc.
We also integrate the Z3 SMT solver to take account of conditional execution and type inheritance.
Other static or dynamic deadlock detectors crashed or gave wrong predictions in some patterns.
Therefore, our type-based deadlock analyzer not only fills the gap in the landscape of value-based detection, but also complements existing detectors.
\end{abstract}


\section{Introduction}
Coroutines, as an abstract programming construct, are a generalization of functions that can suspend execution part-way for later resumption.
To study the collective behavior of a set of coroutines in the field of process calculi, Gu and Ke proposed Coroutine Types~\cite{gu2023typing} which are behavioral type expressions for coroutines.
Their Coroutine Types have been applied to model-driven engineering~\cite{gu2024typing}, smart contracts, and test case generation~\cite{liu2025agts}.

Coroutine Types require a coroutine to fully receive data before yielding anything.
This however becomes a restriction when we apply Coroutine Types to many real-world programming languages that support concurrency.
This paper thus puts forward a Flow extension to Coroutine Types to overcome this restriction.
Our Flow extension allows yielding and receiving operations to happen in any order for an unlimited number of times.
The extension also includes constraint types $\tau / \rho $, akin to the Boolean conditional $\textit{if } \rho \textit{ then } \tau$ in Communicating Sequential Processes~\cite{csp}.
Our constraint types enhance the expressiveness of Coroutine Types so that we are able to take account of type inheritance in data of channels, and any other conditional activation of coroutines.
The reduction rules are adjusted for the new semantics.

To check the applicability and solve real-world problems,
we have created a Flow type system to map expressions in the Go language (abbreviated as \Golang) onto Coroutine Types with Flow.
Thus we are able to check, from the viewpoint of behavioral types, the deadlock freedom of a Go program.
We choose Go because it is a popular choice in building scalable web applications, blockchains, and other distributed systems.
In Go, coroutines are branded as goroutines.

{\Golang} has a wide range of concurrency features~\cite{meyerson2014go},
but our type system only types some of them.
Listing~\ref{allFeatures} shows {\Golang} features that can be mapped onto Coroutine Types with Flow,
including the keyword \code{go} and \code{defer}, and unbuffered channels.
The keyword \code{go} takes a function call or a call to an anonymous subroutine, and starts a goroutine.
A statement prefixed by the keyword \code{defer} defers the execution of a function until the surrounding function returns.

Channels in Go are statically typed conduits, meaning the conduit sends and receives a single type of data.
Meanwhile, one goroutine has access to more than one channel for the input and output purpose.
Channels are created by the syntax \code{make(chan $t$, $n$)}, where $t$ is a type,
and $n$ specifies the buffer capacity of the channel.
If $n$ is 0 or omitted, the channel becomes unbuffered, meaning a send operation will be awaiting until a receive operation is performed on this channel, and vice versa.
The send and the receive operation on a channel use the same operator \code{<-}.
On one hand,
when it's used as a binary operator as in \code{x <- y}, \code{x} is a channel and \code{y} is a value to send.
On the other hand,
when it's used as a unary operator as in \code{<- y}, \code{y} becomes a channel and the whole expression evaluates to the received value.
In Listing~\ref{allFeatures}, Line 4 creates a channel of int;
Line 6 starts a deferred goroutine that reads the channel;
Line 9 sends a datum to the channel.

\begin{lstlisting}[language=go, float=ht, numbers=left, xleftmargin=1.8em,
label=allFeatures, caption={A piece of Go code containing the channel creation, sending, receiving, and the keywords \code{go} and \code{defer}}]
import "fmt"

func main() {
	channel := make(chan int)

	defer func() { fmt.Print(<-channel) }()
	go func() {
		if true {
			channel <- 20
		}
	}()
}
\end{lstlisting}



Coroutine Types with Flow are designed to analyze 17 kinds of deadlock situations with respect to channel and goroutine interactions in \Golang.
We used our algorithm to analyze 4 real-world program files and got correct deadlock situations.
Additionally we ran two state-of-the-art static deadlock analyzers and one dynamic analyzer, and they could not correctly predict deadlock situations in some of the {\TestsCount} cases.

In summary, the main contributions of this paper are:
\begin{enumerate}
\item We propose a Flow extension to the previously published Coroutine Types in order to model flexible receiving and yielding behaviors of coroutines.
\item We propose a set of typing rules that map critical concurrency features in the Go language onto type expressions in the Flow extension.
\item Our Flow extension complements traditional deadlock analyzers which failed to analyze 11 deadlock patterns and 3 real-world cases.
\end{enumerate}

The rest of the paper is organized as follows.
First, we review related work in Section~\ref{sec:Related-Work}.
Then building on previous research, Section~\ref{sec:CoroutineTypes} proposes a Flow extension to Coroutine Types and revises the reduction rules.
Next Section~\ref{sec:type-system} presents a Flow type system that maps {\Golang} expressions to Coroutine Types for detecting deadlocks.
We also compare our algorithm with other tools in this section.
Section~\ref{sec:limitation} briefs about the limitations of Coroutine Types with Flow and the corresponding deadlock detection algorithm.
Lastly we conclude in Section~\ref{sec:conclusion}.

\section{Related Work}\label{sec:Related-Work}

Predicting a deadlock is essentially solving a halting problem, which is proven undecidable~\cite{nielson1994abstract}.
Yet, we can very well reason limited forms of deadlocks~\cite{StaticProgramAnalysis}.
Concurrency bugs have symptoms of either blocking or non-blocking execution~\cite{zhong2022bingo}.
Blocking refers to no progress of all processes, known as a deadlock.
If only a few processes do not make progress, it is a partial lock~\cite{padovani2017type}.

In this section we first talk about abstract interpretation, which is a method to extract certain properties from a program and analyze the properties in an abstract way.
The property in interest can be deadlock-freedom, termination, whether a neural network is robust with respect to perturbations~\cite{singh2019abstract}, so on and so forth.
If we are interested in the type information, abstract interpretation is specialized to a type system,
where a type is not necessarily a standard type, such as integers, booleans, etc., but also behavioral types or session types~\cite{huttel2016foundations} which model of an abstract behavior of a program.
A type system is used in the static time or dynamic time depending on whether the programming language is static or dynamic.
Besides type systems, static or dynamic analysis have other ways to determine concurrency properties of a program.

\subsection{Abstract Interpretation}

Abstract interpretation provides a general framework to approximate program behaviors and derive finite models amenable to automated reasoning.
It reduces the possible states of a program~\cite[p21]{nielson1994abstract}.
One simple example is the analysis of the sign of integer arithmetic, whether the result is positive ($+$), negative ($-$), or unknown ($\pm$).
\etal{Veileborg}~\cite{veileborg2022detecting} abstract the channels in a Go program into status, and use the point-to relation to make a graph of how channels are used.
\cite{laang2019model} translates {\cpp} programs into the modeling language Promela and then finds concurrency issues.

Communicating sequential processes (CSP)~\cite{milner1980calculus} and Calculus of Communicating Systems (CCS)~\cite{csp} are members of process calculi.
CSP supports conditions, and uses an exclamation mark ($!$) for sending data, and an question mark ($?$) for receiving data.
Given {\Golang} source code, Almeida and Antonio~\cite{nova2021} generate an intermediate representation dubbed MiGo~\cite{lange2017fencing}.
MiGo then is translated into a representation in CCS.
They ignored Go's \code{select} statements.
Next, their deadlock detection and resolution algorithms are performed.
Coelho~\cite{coelho2022automatic} modeled \code{select} statements as external choices in CCS and developed a Go module, named GoDDaR.

\subsection{Type Systems}
If types are abstracted from a program, a type system is being studied.
\etal{Hosoya}~\cite{hosoya2005regular} created Regular Expression Types to check the validness of an XML file to a grammar.

Typed Lua~\cite{maidl2014typed} endeavors to introduce optional typing to the Lua language including its coroutines.
While its type system allows for type narrowing, it does not assign a specific type to the coroutine construct as a whole. Furthermore, the return types of coroutines cannot depend on input types within the Typed Lua system.
CorDuroy \cite{anton2011typing} is a statically-typed language for coroutines based on the simply-typed lambda calculus.
Godel \cite{lange2018static} uses behavioral types to model the SSA form of a Go program, and studies the liveness and channel safety properties.
The evaluation was done on a small set of 22 programs.
Godel's method has one limitation in that it does not support programs that spawn new threads in for-loops.
\etal{Barclay}~\cite{barclay2009language} invented a coroutine-based dynamically-typed language for prototyping code.
This language has no functions but coroutines, and all coroutines are callable objects.


\subsection{Static Analysis}


Static tools analyze source code without the need of a compiler or execution of the code.
Static analysis tends to be conservative, in order to avoid the state-explosion problem~\cite{leesatapornwongsa2015samc}.

GCatch~\cite{liu2021automatically} instructs Microsoft Z3~\cite{de2008z3}, a SMT solver to solve an equation that equals to a deadlock state.
If a solution is found, the solution is the conditions that lead to the deadlock.
Staticcheck~\cite{Staticcheck} and Vet~\cite{Vet} analyze Go source code statically. With a limited number of check rules, they spot concurrency bugs in rare cases.
\etal{Zhang} invented a closed-source tool GoDetector~\cite{zhang2021godetector}, which can detect not only channel-related bugs but also WaitGroup-related bugs.
JaDA is a deadlock analyzer for the Java bytecode~\cite{garcia2017jada}.
It was evaluated on 247 Java files in 25 benchmarks.

\subsection{Dynamic Analysis}

Dynamic approaches collect information during execution of a program.
This approach is effective in analyzing one execution of an object program,
but cannot study behaviors which the program has not exhibited.
Dynamic analysis minimizes false positives and negatives, but this method incurs execution overhead.

Memory leaks are often detected at the execution time~\cite{gosain2015survey}.
\etal{Liang}~\cite{liang2016behavior} intercepts API calls from a program to classify whether the program is malicious.

{\Golang} provides the package \code{runtime/trace} \cite{tracePackage}.
Developers place
\begin{center}
\code{trace.Start(traceFile)} and \code{trace.Stop()}
\end{center}
around the code they want to diagnose.
A \code{traceFile} is written to disk, and the command line
\begin{center}
\code{go tool trace traceFile} \cite{traceCommand}
\end{center}
visualizes the file.
This toolchain is not insightful enough to locate concurrency bugs,
so GoAT~\cite{taheri2021goat}, a GOPATH-based project, patches the toolchain so as to record more events, and also devises another static component to visualize and inspect the trace output.
Akin to GoAT, BinGo~\cite{zhong2022bingo} is a hybrid solution involving runtime and static analysis.
Rather than modifying the \code{go} execution binary, its runtime component performs code inject into the address space.
Unfortunately its source code is not available.
Similarly, Concuerror~\cite{christakis2013systematic} combines dynamic analysis with static analysis to find bugs in Erlang.

The most naive deadlock detection resides in the Go runtime itself.
If all goroutines are paused, its built-in deadlock detector will print the stack trace, then kill the program~\cite{ng2016static}.
GoLeak by Uber~\cite{UberGoLeak} declares deadlocks if the main function, which is the root goroutine, fails to finish within a pre-defined time period.

\section{A Flow Extension to Coroutine Types}\label{sec:CoroutineTypes}

The syntax of our behavioral types is defined in Fig.~\ref{fig:coroutine-syntax}.
Italic names---sequence types, product types, and concrete types---are from the base Coroutine Types~\cite{gu2024typing},
not part of the Flow extension.
Concrete types are simple types with no internal structure, such as \tlit{Int} or \tlit{StringBuilder}.
A sequence $\Theta$ is flat and associative, that is $\Seq{\Seq{\tau_1,\tau_2},\tau_3} = \Seq{\tau_1,\Seq{\tau_2,\tau_3}}$, where both sides can be simplified to $\Seq{\tau_1,\tau_2,\tau_3}$.
A sequence of a particular type $\tau$ is written as $\Theta_\tau$.
To give the length $n$ of a sequence, $\Theta_\tau$ is written as $\tau^n$.

\begin{figure}[ht]
\begin{displaymath}
\begin{aligned}
\omega ::=&  & \mbox{Flow item} &\ \Omega \\
\mid &\ \Yield{\tau}  &\mbox{yielded item} \\
\mid &\ \Receive{\tau}  &\mbox{received item}\\
\tau ::=& 					&\mbox{types} &\ T \\
\mid &\ \tau \slash \rho		&\mbox{constrained types} &\ \Xi \\
\mid &\ (\tau \mid \tau)  		& \mbox{union types} &\ \Sigma \\
\mid &\ \corD{{\omega}{\omega}{\cdots}} & \mbox{coroutine def types} &\ D \\
\mid &\ \corI{{\omega}{\omega}{\cdots}} & \mbox{coroutine ins types} &\ I \\
\mid &\ \Seq{\tau,\tau}  		& \mbox{\textit{sequence types}} &\ \Theta \\
\mid &\ \Tuple{\tau,\tau} 		& \mbox{\textit{product types}} &\ P \\
\mid &\ \tlit{Int} \mid \tlit{String} \mid \cdots & \mbox{\textit{concrete types}} &\ K \\
\end{aligned}
\end{displaymath}
\caption{Abstract syntax for our coroutine types with Flow}
\label{fig:coroutine-syntax}
\end{figure}

The principal types in the Flow extension are coroutine definition types $D$ and coroutine instance types $I$.
A coroutine definition is akin to a function definition.
A coroutine instance is one execution of a function. Depending on the arguments, coroutine instances may exhibit diverse input and output behaviors.
$\Start: D \to I$ is a type-level function that starts a coroutine definition and turns it into a coroutine instance.
Both definition and instance are made of Flow items $\Theta_\Omega$, telling how the coroutine yields and receives data.
Yielding and receiving actions are a protocol to control when to execute a coroutine.
Self-canceling is permitted, for example $\corI{{\Yield{A}}{\Receive{A}}}$.
Disjoint union types may exist in a coroutine definition
but not in a coroutine instance because at runtime we always know the exact chosen type.


A type can take an optional Boolean predicate $\rho$.
The instructions in $\rho$ are specific to applications.
It could model type inheritance, generic functions, and conditional executions.
For instance,
$$
l: \corD{{\Receive{\Seq{x,\tlit{Book}}}}{\Yield{\Seq{x,\tlit{Book},\tlit{Reserve}}}}} \slash x \mathop{<:}  \tlit{User}
$$
reads: ``$l$ is a coroutine definition which receives $x$ and \tlit{Book}, and yields $x$, \tlit{Book}, and \tlit{Reserve}, where $x$ is a subclass of \tlit{User}.''
The kinds of predicates depend on the application and is not part of the Flow extension.
Equation \eqref{eq:constraints} is reduction rules of constrained types.
The construct $\Zero$ denotes no behavior.

\begin{equation}\label{eq:constraints}
\begin{aligned}
\tau \slash \tlit{false} &\Rightarrow \Zero \\
\tau \slash \tlit{true} &\Rightarrow \tau \\
\tau \slash \rho_1 \slash \rho_2 &\Rightarrow \tau \slash \rho_1 \wedge \rho_2 \\
\tau \slash \rho \wedge (x \mapsto v) &\Rightarrow \tau [x \mapsto v] \slash \rho
\end{aligned}
\end{equation}

If the type in a Flow item is a sequence, the Flow direction indicator $\psi$ (yielding $!$ or receiving $?$) should be distributed to each sequence item,
i.e., ${\psi}{\Seq{\tau_1,\tau_2, \cdots}} \Rightarrow \Seq{{\psi}{\tau_1},{\psi}{\tau_2}, \cdots}$.

Variables can take place of concrete types, integers, or arguments of a function.
Free variables are considered constants.

Gu and Ke's Coroutine Types~\cite{gu2023typing} defines minimal elements for coroutine reduction,
containing solely concrete types, sequences, coroutines with fixed receiving and yielding order.
Coroutine definitions and coroutine instances are not distinguished there.
They later added product types to control the precedence of reduction~\cite{gu2024typing},
but here we switch to leverage $\Start$ to activate coroutines, which is to some extent a precedence.
We still keep product types as tuples because they can group elements.




\subsection{Reduction Rules}

The reduce function is defined as $\compose{}:\Theta_I \to I$.
The arguments of $\compose{}$ have to be a sequence of coroutine instances or Start functions that return instances.
The return type is a coroutine instance $I$ and we do not further reduce it to $K$.
In this way we know the data are computed rather than from a static field.
The reduce function $\compose$ consists of rules running in a demand-driven strategy~\cite{papazoglou1984outline}.

During the evaluation of $\compose$,
a Flow item with empty type is regarded as useless, and should be removed from a coroutine.
$\psi$ being the channel operation $!$ or $?$,
reduction Rule~\ref{eq:remove-void} lists the reduction regarding removing these identity elements.
$e_1 \Rightarrow e_2$ means an expression $e_1$ evaluates or reduces to another expression or value $e_2$.

\begin{figure*}
\begin{composingEquation}\label{eq:remove-void}
\begin{split}
\compose\Seq{S_1,\corI{{\omega_1}{\psi(\Zero)}{\omega_2}},S_2} &\Rightarrow \compose\Seq{S_1,\corI{{\omega_1}{\omega_2}},S_2} \\
\compose\Seq{S_1,\corI{{\psi(\Zero)}},S_2} &\Rightarrow \compose\Seq{S_1,S_2} \\
\compose\Seq{S_1,\Start(\corD{{\omega_1}{\psi(\Zero)}{\omega_2}}),S_2} &\Rightarrow \compose\Seq{S_1,\Start(\corD{{\omega_1}{\omega_2}}),S_2} \\
\compose\Seq{S_1,\Start(\corD{{\psi(\Zero)}}),S_2} &\Rightarrow \compose\Seq{S_1,S_2} \\
\end{split}
\end{composingEquation}
\end{figure*}

$\Start(d)$ takes a coroutine definition $d$ and turns it into a coroutine instance $i$.
Recall a coroutine definition may contain disjoint union types.
In Listing~\ref{go-conditional},
if we ignore the \code{defer} statement,
we have $s: \corD{{\Yield{(\tlit{Int} / v < 10 \mid \tlit{Bool} / \neg (v < 10))}}} $ for the definition.
For instantiation, \code{go s(20)} translates $\Start(s)$ to $\corI{{\Yield{\tlit{Bool}}}}$
because the branch $\tlit{Bool}  / \neg (v < 10)$ is chosen,
while \code{go s(2)} translates  $\Start(s)$ to $\corI{{\Yield{\tlit{Int}}}}$.
Section~\ref{sec:type-system} will discuss the mapping from the if statement to $(\tlit{Int} / v < 10 \mid \tlit{Bool} / \neg (v < 10))$,
as well as the processing of Go's \code{defer} statements.

\begin{lstlisting}[language=go, float=ht, label=go-conditional, caption={Function s yields disparate types based on a condition, which can be solved by the Z3 solver}]
var ch1 chan int = make(chan int)
var ch2 chan bool = make(chan bool)

func s(v int) {
	defer func(){ ch2 <- true }()
	if v < 10 { ch1 <- v
	} else { ch2 <- true }
}

func main() {
	go s(2)

	fmt.Println(<- ch1, <- ch2)
}
\end{lstlisting}

Other reduction rules need to access or manipulate auxiliary data.
Thus the symbol $\vdash$ is employed. Left to $\vdash$ is the context comprised of Pending Type $\tau: T$ and External Yields $E: \Theta_{K \mid P}$.
Pending type is what a coroutine yielded and not yet received by a coroutine.
External Yields is a sequence of pending types that are not receivable by any coroutine.
Right to $\vdash$ is an expression which can be a type or a call to the reduce function.
Same as the previous work~\cite{gu2024typing},
we use a wildcard $(\cdot)$ to indicate an unreferenced item in rules. For example $(\cdot,E)$ means at this point the pending type, which may or may not be $\Zero$, is not used in other parts of the rule.

The ``main body'' of coroutine instance and coroutine definition is a sequence of Flow items.
We use the head function and the tail function, defined in the base Coroutine Types~\cite{gu2024typing}
to get the first element and rest elements of an instance, shown in Fig.~\ref{head-tail}.
We never take head or tail out of a definition.

\begin{figure}
\begin{displaymath}
\begin{aligned}
\Hd(\corI{{\omega_1}{\omega_2}{\cdots}}) &= \omega_1 \\
\Tl(\corI{{\omega_1}{\omega_2}{\cdots}}) &= \corI{{\omega_2}{\cdots}} \\
\end{aligned}
\end{displaymath}
\caption{Definition of head $\Hd$ and tail $\Tl$ of a coroutine instance}\label{head-tail}
\end{figure}

The following reduction rules \eqref{eq:yield} to \eqref{eq:resume-co} are very similar to the ones in the base Coroutine Types,
but we have to revise them to recognize the new Flow items $\Omega$.

\subsubsection{Yielding}

\begin{lstlisting}[language=haskell, float=ht, label=first-in-haskel, caption={the function $\First$ and $\NotFound$ in Haskell}]
first :: [a] -> (a -> Bool) -> (Maybe a, [a], [a])
first s p =
  case break p s of
    (before, [])      -> (Nothing, before, [])
    (before, x:after) -> (Just x, before, after)

none :: [a] -> (a -> Bool) -> Bool
none s p =
  case first s p of
    (Nothing, s, []) -> True
    _ -> False
\end{lstlisting}

We use the function $\First$ defined in the base Coroutine Types to find the first coroutine instance $i$ in a list $S$ that matches a condition $p$, written as
$(i, S_1, S_2) = \First(S, \lambda i.p(i))$,
where along with the coroutine, $\First$ also returns $S_1$ all elements before $i$ and $S_2$ all elements after $i$.
Listing~\ref{first-in-haskel} is its formal definition in Haskell.
If $\First$ can't make out a matching element, then $i = S_2 = \Zero$.
This special case is abbreviated as $\NotFound$.

\eqref{eq:yield} is triggered when there is no pending type and
the function $\First$ is able to find the a coroutine whose head direction is yielding.
Then, we transfer the yielded type into the context.
In case a coroutine has exhausted its action statements, $\Tl(s)$ would be $\Zero$ and it's subject to deletion by \eqref{eq:remove-void}.

\begin{composingEquation}
\begin{split}
(\Zero, \cdot) \vdash \compose(S) \Rightarrow
  (\tau, \cdot ) \vdash \compose(\Seq{S_1, {\Tl(i)}, S_2})
   \\\Provided\; \\
(i,S_1, S_2) = \First(S, \lambda i\ldotp \Hd(i)=\Yield{\tau})
\end{split}
\label{eq:yield}
\end{composingEquation}

If the yielded type is a coroutine instance or an invocation to the type-level function $\Start$,
\eqref{eq:yield-co} evaluates $\Start$ and inserts the yielded instance at the end of the list.
This mimics the breadth-first search and mitigates infinite loops if a coroutine continuously starts itself.

\begin{composingEquation}
\begin{split}
(\Zero, \cdot )\vdash \compose(S) \Rightarrow
  (\Zero, \cdot) \vdash \compose(\Seq{S_1, {\Tl(i)}, S_2, \corI{{u}{v}}})
  \\\Provided \quad\\
(i,S_1, S_2) = \First(S, \lambda i\ldotp \Hd(s)=\Yield{\corI{{u}{v}}})
\end{split}
\label{eq:yield-co}
\end{composingEquation}

\eqref{eq:co-to-ext} describes a deadlocked state when no coroutine has a yielding head and the pending type is $\Zero$, suggesting nothing to receive.
In such case, the reduction terminates with result $E$ followed by all the remaining coroutines from the list.
The distributive law applies in this rule, and coroutines as a first-class citizen are allowed within a coroutine.

\begin{composingEquation}
\begin{split}
(\Zero, E) \vdash \compose(S) \Rightarrow
  (\Zero, \Zero) \vdash \corI{{\Yield{E}}{\Yield{S}}} \\\Provided\;
\NotFound(S, \lambda i\ldotp \Hd(i)=\Yield{\tau})
\end{split}
\label{eq:co-to-ext}
\end{composingEquation}

%
%

\subsubsection{Resuming}

When the pending type $\tau$ is not $\Zero$, the resume operation is triggered.
\eqref{eq:resume} picks up the first coroutine $i$ that can receive $\tau$, and resumes it.
$\Match: T\times T \to \Theta$
is defined in the base Coroutine Types,
a commutative function to match two types (potentially carrying constraints) and returns a set of conditions $B$
under which the two types are equal.
As we know, a coroutine type can contain variables and therefore the conditions returned by $\Match$ could bind the variables.
For example, $\Match(\tlit{Int}^5,\tlit{Int}^n)=\{n \mapsto 5\}$. If in no way can two types match, $\bot$ is returned,
and $\bot \wedge B = \bot$.
Our improvement to the $\Match$ function is to back it up by a SMT solver, Microsoft Z3.
The details about constraint handling are given in Section~\ref{sec:z3}.

\begin{figure*}
\begin{composingEquation}
\begin{split}
(\tau, \cdot) \vdash \compose(S) \Rightarrow
  (\Zero, \cdot) \vdash \compose\Seq{S_1, \Tl(i) \slash B, S_2 }  \quad \Provided \;\\
(i,S_1, S_2) = \First(S, \lambda i\ldotp \Hd(i) = \Receive{s} \wedge \Match(\tau,s)=B)
\end{split}
\label{eq:resume}
\end{composingEquation}
\begin{composingEquation}
\begin{split}
(\tau ,E)\vdash \compose(S)  \Rightarrow (\Zero, \Seq{E,\tau}) \vdash \compose(S) \quad \Provided \;\\
\NotFound(S, \lambda i\ldotp \Hd(i)=\Receive{u} \wedge \Match(\tau,u) \neq \bot)
\end{split}
\label{eq:external}
\end{composingEquation}
\begin{composingEquation}
\begin{split}
(\Zero, \cdot )\vdash \compose(S)  \Rightarrow (\Zero, \cdot ) \vdash \compose(\Seq{S_1, {\Tl(i)} \slash B, S_2} \smallsetminus i') \quad
\Provided\;\\
(i,S_1, S_2) = \First(S,\lambda i\ldotp \Hd(i)=\Receive{t} \wedge t\in I ) \\
\textrm{and}\;
(i', \cdot, \cdot) = \First(S \smallsetminus i,\lambda i'\ldotp \Match(i', t)=B)
\end{split}
\label{eq:resume-co}
\end{composingEquation}
\end{figure*}

\eqref{eq:external} stipulates that if none of the coroutines in $S$ can process the pending type $\tau$, then $\tau$ gets appended to the external yields $E$.
This reduction rule often paves the way for \eqref{eq:co-to-ext} for termination
because \eqref{eq:co-to-ext} expects an empty pending type.

Coroutine cancelation is not a key part for {\Golang} because goroutine must be written with a semaphore or alike to allow early termination.
Nevertheless, \eqref{eq:resume-co} outlines how a coroutine receives other coroutines
for the complete first-class support of coroutine in the base Coroutine Types.
Line 2 in \eqref{eq:resume-co} finds the first coroutine whose head $\Hd(s)$ is a receiving pattern of a coroutine instance,
then Line 3 matches this pattern with another coroutine $i'$, returning conditions $B$.
At the end, we apply the condition to $\Tl(i)$, and also remove the received coroutine from the result by using the set minus sign.
Overall this rule shows the powerfulness of the first-class coroutine support: a component of a coroutine can remove another coroutine as a whole.

The base Coroutine Types~\cite{gu2024typing} has a rule that loops the external yields. This rule is detrimental to channel-based deadlock detection.
Consequently our Flow extension disables this rule.
Consider Listing~\ref{out-of-order}, where \code{main()} first yields String, then yields Int, but \code{work()} requires Int followed by String.
The order mismatch makes for infinite blocking. If we enable the rule ``loop external'' in \cite{gu2024typing}, the analyzer will report termination, which is incorrect.
Listing~\ref{out-of-order} was enlisted into the supported deadlock patterns in Section~\ref{sec:evaluation}.

\begin{lstlisting}[language=go, float=ht, label=out-of-order, caption={\code{Out-of-order.go} in our dataset will run into a deadlock}]
func work(cInt chan int, cStr chan string) {
	fmt.Println(<-cInt)
	fmt.Println(<-cStr)
}

func main() {
	cInt := make(chan int)
	cStr := make(chan string)
	go work(cInt, cStr)

	cStr <- "hello"
	cInt <- 1

	...
}
\end{lstlisting}

Overall, the whole reduction algorithm is deterministic because it checks rules in order and there is no random choice.
The step to collect coroutine definitions surely terminates because there is a limited number of functions in a program.
Nevertheless, the subsequent reduction process may not terminate if a Go function recursively calls itself.
This is one form of labeled infinite coroutines that always restore itself after execution.
\eqref{eq:yield-co} adopts the breath-first manner and mitigates infinite loops.
Moreover, the whole reduction process aborts once the main coroutine is used up.
As the last measure, our implementation caps the reduction steps at 500.
\ifblind Gu and Ke~\cite{gu2023typing} had \else Our previous paper~\cite{gu2023typing} has\fi detailed analysis of these properties, including determinism and termination.

\subsection{Constrained Type Handling}\label{sec:z3}
Whenever $\Match$ is called, it is possible that a constrained type needs to be checked.
This kind of checking is done by Microsoft Z3~\cite{de2008z3} in the Flow extension.
Z3 recognizes many logical operations, such as negation, conjunction and disjunction.
Equality checks on symbols and comparisons (less-than and greater-than) on numbers are also fundamental.
Of these binary relations, on one side is typically a variable, and on the other side is a constant or another variable.
In our case, symbols are our concrete types in Fig.~\ref{fig:coroutine-syntax}.

\begin{figure}
\begin{displaymath}
\begin{aligned}
c(\corI{{\omega_1}{\omega_2}{\cdots}{\omega_n}}) &= \textstyle\bigcup c(\omega_i) \\
c(\corD{{\omega_1}{\omega_2}{\cdots}{\omega_n}}) &= \textstyle\bigcup c(\omega_i) \\
c(\psi\tau) &= c(\tau) \\
c(\tau \slash \rho) &= c(\tau) \cup c(\rho) \\
c(\tau_1 \mid \tau_2) &= c(\tau_1) \cup c(\tau_2) \\
c(\Seq{\tau_1,\tau_2,\cdots,\tau_n}) &= \textstyle\bigcup c(\tau_i) \\
c(\Tuple{\tau_1,\tau_2,\cdots,\tau_n}) &= \textstyle\bigcup c(\tau_i) \\
c(K) &= \left\{ K \right\}
\end{aligned}
\end{displaymath}
\caption{Function $c$ finds all concrete types from a type}
\label{fig:collect}
\end{figure}

\begin{figure}[ht]
\centering
\includegraphics[width=0.9\linewidth]{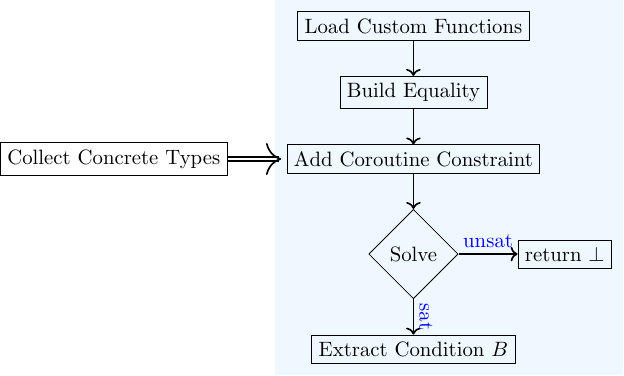}
\caption{Collect Concrete Types runs only once. The shaded box on the right is the flowchart of the $\Match$ function}
\label{fig:z3-flowchart}
\end{figure}

As mentioned, variables in our coroutine types are allowed to present any concrete type.
The set containing all concrete types in the universe is named the \code{Concrete} sort in Z3.
We denote the \code{Concrete} sort as $\mathbf{K}$.
Let $c: T \to \Theta_K$ be a function that collects all concrete types from a type.
Fig.~\ref{fig:collect} shows the definition of $c$.
The collect function runs before reduction rules introduced earlier because we need to collect all kinds of types in the program.
The assembled \code{Concrete} sort $\mathbf{K}$ is fed into the $\Match$ function highlighted by the shaded background on the right of Fig.~\ref{fig:z3-flowchart}.

$\Match$ (and $\NotFound$) involves in building a boolean expression,
solving the expression, and returning the condition under which the expression is true.
In the Flow extension, the types of variables are inferred.
If the opposite side of a variable is an integer, the variable is of type Int.
If opposite a variable is a concrete type, the variable is of type Concrete and given the set of collected concrete types as its domain.
If both sides are variables, the two should have the same type.
A variable equal to a complex type or $\Zero$ is not permitted, resulting to $\bot$.
$\Match$ processes not only equality checks, but also arithmetic comparisons or custom functions that may appear in $\rho$ in constrained types.
Examples include
the less-than symbol in $\tlit{Int} \slash v<10$,
and the subtype constraint in $\big [{\Receive{\Seq{x,\tlit{Book}}}};\linebreak[1]{\Yield{\Seq{x,\tlit{Book},\tlit{Reserve}}}} \big ] \slash x \mathop{<:} \tlit{User}$

Although Z3 comes with assorted Boolean and numerical expressions,
users of the Flow extension, such as the Flow type system introduced in Section~\ref{sec:type-system},
can inject custom functions, such as a subtype relation, for application-specific purposes.
Custom functions shall return a Boolean value.
They are usually second-order logic with universal qualifiers and a material conditional.
In a Go program, type inheritance is usually achieved by anonymous fields.
We define a function head \code{inherit} and a function body in SMT-LIB syntax~\cite{barrett2010smt}.
Listing~\ref{inherit-body} establishes a function body, accepting \code{x} of \code{Concrete} and \code{y} of \code{Concrete}.
\code{r} is the inverse of known subtyping relations, such as Faculty is a subclass of User, and Student is a subclass of User.
Then \code{=>} is an implication (material conditional) that if \code{x} and \code{y} are not in known subtyping relationships, \code{(inherit x y)} is false.

We could have kept a persistent Z3 context and run Load Custom Functions once as we do for Collect Concrete Types,
but Z3 may have some memory leaking issue so we have to free up the Z3 context regularly.

\begin{lstlisting}[columns=fixed, language={}, morekeywords={forall,let,not,or,and, false}, otherkeywords={=>},
float, label=inherit-body, caption={Function body of the inherit custom function}]
(forall ((x Concrete) (y Concrete))
  (let ((r (not (or (and (= x Faculty) (= y User))
                    (and (= x Student) (= y User))))))
    (=> r (= (inherit x y) false))))
\end{lstlisting}

When $\Match$ is called with a pending type $\tau_1 / \rho_1$ and a receiving coroutine $\corI{{\Receive{\tau_2}}{\cdots}} \slash \rho_2$,
we build an Boolean expression $(\tau_1 = \tau_2) \wedge \rho_1 \wedge \rho_2$, and ask Z3 to solve it.
Equation \eqref{eq:constraints} is used as well.
If Z3 answers ``sat'' on the Boolean expression, we extract condition $B$ from the interpretation
(the step Extract Condition) and use it in \eqref{eq:resume} or \eqref{eq:resume-co}.

Z3 is a satisfiability solver and only finds one instance of variables that make the expression true.
Suppose during \eqref{eq:resume}, the pending type is $\Tuple{x^1,y^j} \slash j<5$,
and $\Match$ is going to process the head of $\corI{{\Receive{\Tuple{x^i,y^j}}}{\Receive{\Tuple{y^j,x^i}}}} \slash j>0$,
and $i,j$ are variables;
then we build an expression $ \Tuple{x^1,y^j}= \Tuple{x^i,y^j} \wedge j<5 \wedge j>0 $.
Z3 will return $B=\left\{ i \mapsto 1, j \mapsto 1 \right\}$.
While $j=1$ satisfies the expression,
we shouldn't replace $j$ by this exact number because the value of $j$ is not uniquely determined yet.
Instead, we should update the coroutine to $\corI{{{-}{\Tuple{y^j,x^1}}}} \slash 0<j<5$.

To check if a variable interpretation is unique or not, we append the inverse of the interpretation after the initial solving, and check again.
If the check passes, the interpretation is not unique and should be excluded from the return condition.

\section{A Flow Type System for Goroutine Deadlocks}\label{sec:type-system}

{\Golang} starts the main function as a root goroutine~\cite[p9]{coelho2022automatic}, and it can spawn new goroutines.
If all goroutines are in the waiting state---a complete halt of the program's progress, the go runtime believes the program is never-ending,
and reports ``fatal error: all goroutines are asleep - deadlock!''~\cite{ng2016static}


We apply the newly created Flow extension to detect deadlocks in specific patterns.
By and large, we map the keyword \code{go} onto the $\Start$ function;
function definitions onto coroutine definitions;
send and receive operations of a channel onto our yielding and receiving primitives.

To begin with, we parse the source code and learn function definitions.
Equation \eqref{eq:cordef-iteration} checks if a function in {\Golang} is our coroutine function.
If a Go function taps into channels to send or receive data,
a coroutine definition is created for this function,
with the send operation typed as $\Yield{\tau}$ and the receive operation typed as $\Receive{\tau}$.
If a function calls another coroutine $f'$, this function is a coroutine definition with a Flow item $\Inline(f')$.
If a function starts another coroutine $f'$ with keyword \code{go}, this function is a coroutine too with a Flow item $\Start(f')$.
Otherwise this function is not in the range of $M'$.
Then, Equation \eqref{eq:cordef-fix} defines the function (or mapping) $M$ by evaluating the fixed point of $M'$.

\begin{equation}\label{eq:cordef-iteration}
M'(f) = \begin{cases}
\corD{{\cdot}{\cdot}} & \text{if $f$ uses channels} \\
\corD{{\cdot}{\cdot}} & \text{if $f$ calls } ~ f' \in \operatorname{Range}(M') \\
\Zero & \text{otherwise.}
\end{cases}
\end{equation}

\begin{equation}\label{eq:cordef-fix}
M = \operatorname{fix}(M')
\end{equation}

Function $\Inline$ is like $\Start$ that turns a coroutine definition into a coroutine instance,
but it adds a new coroutine instance into the current Flow items rather than the reduction sequence.
Keyword \code{defer} is treated as $\Inline$ but the operations are inserted at the end of the coroutine Flow.


Having typed Go functions as coroutine definition types, we commence deadlock checking by investigating the main function.
The external yields is disabled because we are analyzing an executable rather than a library, and we require channel sizes are 0.
This main function is also short-circuited considering that once the main function finishes, the Go runtime will kill all remaining goroutines.

To start a coroutine definition, we conduct Code Flow Analysis (CFA)~\cite{CFA} to determine values and types of arguments.
If CFA cannot statically determine the truth value of an if-conditional expression,
we map the conditional expression to the $\rho$ part of the Flow extension, which will be solved by Z3.
Formally, \code{if $p$ \{$s_1$; $s_2$\} else \{$s_3$;$s_4$\}} is mapped to
$\left( \Seq{s_1,s_2} / p==\tlit{true}  \mid \Seq{s_3,s_4} / p==\tlit{false} \right)$.
To illustrate, see Listing~\ref{conditional}.
We get
$\corD{{\Yield{\tlit{Int}}}} / 1 \leqslant \mathit{weekday} \leqslant 3 $ and
$\corD{{\Receive{\tlit{Int}}}} / 3 \leqslant \mathit{weekday} \leqslant 5 $.
Z3 can divide the value of $\mathit{weekday}$ into 4 groups, 1 to 2, 3, 4 to 5, and 6 to 7.
We then perform a reduction for each group. Apparently, on Wednesday (3) the sending and receiving operation are paired, resulting in no lock.
At weekends (6 to 7), there are no sending and receiving, and there is no concurrency issue.
On other days, the program suffers from deadlocks.

\begin{lstlisting}[language=go, float=ht, label=conditional, caption={Conditional goroutines are mapped onto $\tau / \rho$}]
func main() {
	...

	ch := make(chan int)
	if 1 <= weekday && weekday <= 3 {
		go func() {
			ch <- 1
		}()
	}

	if 3 <= weekday && weekday <= 5 {
		go func() {
			fmt.Println(<-ch)
		}()
	}

	...
}
\end{lstlisting}

We use Listing~\ref{moby4395} to illustrate the whole workflow.
This snippet defines a \code{main} function and a \code{run} function.
After two iterations, $M$ reaches the fixed point.
Function \code{run} starts an anonymous goroutine, so $ M(\mathit{run})=\corD{{\Start(\corD{{\Yield{\tlit{Error}}}})}} $.
Function \code{main} calls \code{run()}, so the first Flow item is $\Inline(\mathit{run})$,
then it receives a datum from channel of type Error, so the next Flow item is $\Receive\tlit{Error}$.

\begin{lstlisting}[language=go, deletekeywords={error}, float=ht, label=moby4395, caption={A deadlock-free version of moby\#4395, where
$M(\mathit{main}) = \protect\corD{{\Inline(\mathit{run})}{\Receive{\tlit{Error}}}}$}]
func run(f func() Error) chan Error {
	ch := make(chan Error)
	go func() {
		ch <- f()
	}()
	return ch
}

func main() {
	err := run(func() Error {
		return nil
	})

	<-err
}
\end{lstlisting}

\begin{figure}[ht]
\begin{align*}
& \compose \Seq{\Start(\mathit{main})} \\
\Rightarrow & \compose\Seq{\corI{{\Inline(\mathit{run})}{\Receive\tlit{Error}}}}  \\
\Rightarrow & \compose\Seq{\corI{{\corI{{\Start(\corD{{\Yield\tlit{Error}}})}}}{\Receive\tlit{Error}}}} \\
\Rightarrow & \compose\Seq{\corI{{\Receive\tlit{Error}}}, \corI{{\Yield\tlit{Error}}}} \\
\Rightarrow & ~ \tlit{Error} \vdash \compose\Seq{\corI{{\Receive\tlit{Error}}}} \\
\Rightarrow & ~ \Zero
\end{align*}
\caption{Reduction process of moby\#4395}
\label{fig:composition-mody4395}
\end{figure}

Fig.~\ref{fig:composition-mody4395} shows the reduction process of moby\#4395.
$\Inline$ adds the Flow item of $\mathit{run}$ into the current coroutine instance,
then $\Start$ creates a new coroutine, so $\compose()$ has two arguments now.
Accordingly, \eqref{eq:yield} is activated and puts \tlit{Error} into the context.
Finally \eqref{eq:resume} and \eqref{eq:remove-void} run and return the reduction result $\Zero$.
Since all types cancel out, this piece of code is synchronized.

A deadlocked example is Listing~\ref{out-of-order}.
We have $M(\mathit{work})=\corD{{\Receive\tlit{Int}}{\Receive\tlit{String}}}$,
$M(\mathit{main})=\corD{{\Start(\mathit{work})}{\Yield{\tlit{String}}}{\Yield{\tlit{Int}}}}$.
The reduction result is $\corI{{\Yield{\tlit{String}}}}$, which means this piece of code will yield String to its caller,
but since this is an executable with no caller, the program will hang.

\subsection{Comparison with Other Deadlock Analyzers}\label{sec:evaluation}

We apply Coroutine Types with Flow to the deadlock detection problem for Go programs.
For one channel, its sending site can be at the method body, inside a regular function, inside a Start call, or inside a Defer call.
There are 4 situations.
The receiving site also has 4 situations.
From the $4 \times 4 = 16$ situations we can create a case with locks and a case without locks.
Therefore, a single channel with a pair of sending and yielding has 32 usage patterns not yet considering nested calls.

Rather than evaluating all 32 cases,
we collected 17 single-file {\Golang} programs from \cite{YumaInaura}, \cite{go101}, and other sources, presenting typical usages of channels and goroutines.
Among them 10 are free from deadlocks, and the rest 7 have deadlocks.
Some use more than one channel, and some has nested function calls.
Our algorithm has been designed to recognize these 17 patterns of interactions, listed in Table~\ref{deadlockPatterns}.

\newcounter{pattern}
\renewcommand{\thepattern}{P\arabic{pattern}}

\newcommand{\pattern}[1]{%
  \refstepcounter{pattern}%
  \label{#1}%
  \thepattern
}

\begin{table*}
\centering
\caption{The descriptions of 17 interaction patterns of channels and goroutines}
\label{deadlockPatterns}
\begin{tabular}{llll}
Pattern \# & Short Name                        & Deadlocks & Description    \\\hline
\pattern{pt:basic} & basic                         & No           & Calculate the sum of two arrays. Send and receive the results in two channels \\
\pattern{pt:basic-receiveInFunction} & basic-receiveInFunction       & No           & A variant of P1, with the receiving in a different function\\
\pattern{pt:defer2} & defer2                        & No           & Nested \code{defer} calls \\
\pattern{pt:inline-func} & inline-func                   & No           & Anonymous functions with goroutines and channel communication     \\
\pattern{pt:inline-func-varOutside} & inline-func-varOutside        & No           & Anonymous functions that captures a variable from outside scope   \\
\pattern{pt:main-exit} & main-exit                     & No           & The main function finishes before goroutines complete \\
\pattern{pt:rec-main} & rec-main                      & No           & Recursive goroutines \\
\pattern{pt:SleepingReceiver} & SleepingReceiver              & No           & A receiver goroutine waits on a channel\\
\pattern{pt:SleepingSender} & SleepingSender                & No           & A sender goroutine waits and then sends data \\
\pattern{pt:wait30} & wait30                        & No           & Sleeps for certain time and then exists \\
\pattern{pt:basic-receiveInFunction-extra} & basic-receiveInFunction-extra & Yes          & A variant of P1, with the receiving in a different function, but mis-paired.\\
\pattern{pt:basic3receive} & basic3receive                 & Yes           & Attempt to receive more data than sent, causing a deadlock \\
\pattern{pt:NoSender} & NoSender                      & Yes          & A receiving operation is attempted, but no data are sent\\
\pattern{pt:NoReceiver} & NoReceiver                      & Yes          & A sending operation is attempted, but no data are received\\
\pattern{pt:NoLiveGoroutines} & NoLiveGoroutines              & Yes          & A variant of P13 with two goroutines \\
\pattern{pt:func} & func                          & Yes          & A variant of P13 with two channels \\
\pattern{pt:out-of-order} & out-of-order                  & Yes          & Two channels of different types are used out of order\\
\pattern{pt:return-channel} & return-channel                & Yes          & A channel is returned from a function and receiving and sending are not paired
\end{tabular}
\end{table*}

\etal{Yuan}~\cite{yuan2021gobench} sampled famous open-source repositories and extracted 68 blocking bugs.
Among 68 bugs, only only 4 bugs are exclusively due to unbuffered channels and send/receive primitives.
Thus we used the 4 files to test the applicability of our algorithm to real-world programs.
Our coroutine reduction algorithm could successfully identify these 4 cases.
The 17 artificial cases and 4 real-world cases are listed in Table~\ref{go-test-cases}.
We released the {\TestsCount} test files and the source code of the Flow extension for readers to reproduce~\cite{source-code-anonymous}.

We picked two static tools and one dynamic tool to check their deadlock detection accuracy.
In Table~\ref{go-test-cases}, if a file has global deadlocks, it's labeled as Yes, otherwise No.
Incorrect verdicts are made bold.
In \ref{pt:main-exit}, the main thread exits before other goroutines.
GOMELA treats it as a deadlock, GoAT treats it a partial deadlock (PDL).
This is relevant to subtleties of deadlocks, and we do not mark these verdicts wrong.
All these tools were ran under their default settings.
Our algorithm has no parameter to fine-tune.

GoDDaR v2.0.2~\cite{coelho2022automatic} errored out for 8 files.
For \code{moby\#33293.go}, GoDDaR throws the error ``Invalid argument''.
For other files, this tool throws the error ``Recursive call'' even though some of these files do not have recursive calls.
We suspect GoDDaR has programming issues in recognizing our test cases.

Being a dynamic tool, GoAT~\cite{taheri2021goat} is almost correct for all cases.
However, the Achilles' heel of GoAT is its timeout setting.
If a subject needs more time than the pre-defined timeout to finish execution, GoAT prematurely asserts a deadlock.
That's why GoAT labeled \ref{pt:wait30} deadlocked.
GoAT also had to spend up 30 seconds before giving all GDL verdicts, while the other tools gave verdicts in a flash.

GOMELA-ase21~\cite{dilley2022bounded}, an improved version of GoDel,
is not shy about crashing due to its limited syntax support.
The error messages include
``Function declared as a variable'',
``A receive was found on a channel'',
``Returning a channel'', etc.
We are not sure why GOMELA did not allow a receive operation on a channel.
Also GOMELA outputs nothing for \ref{pt:wait30}.

\begin{table}
\centering
\caption{Analysis results of our algorithm and other 3 tools}
\label{go-test-cases}
\begin{tabular}{lllll}
Program               & Ours & GoDDaR       & GOMELA       & GoAT         \\ \hline
\ref{pt:basic}                         & No      & \textbf{Err} & No           & No           \\
\ref{pt:basic-receiveInFunction}       & No      & \textbf{Err} & No           & No           \\
\ref{pt:defer2}                        & No	   & \textbf{Yes} & \textbf{Err} & No           \\
\ref{pt:inline-func}                   & No      & No           & \textbf{Err} & No           \\
\ref{pt:inline-func-varOutside}        & No      & No           & \textbf{Err} & No           \\
\ref{pt:main-exit}                     & No      & \textbf{Err} & Yes          & PDL          \\
\ref{pt:rec-main}                      & No      & \textbf{Yes} & \textbf{Err} & PDL          \\
\ref{pt:SleepingReceiver}              & No      & No           & No           & No           \\
\ref{pt:SleepingSender}                & No      & No           & No           & No           \\
\ref{pt:wait30}                        & No      & No           & \textbf{Err} & \textbf{Yes} \\
\ref{pt:basic-receiveInFunction-extra} & Yes     & \textbf{Err} & Yes          & Yes          \\
\ref{pt:basic3receive}                 & Yes     & \textbf{Err} & Yes          & Yes          \\
\ref{pt:func}                          & Yes     & Yes          & Yes          & Yes          \\
\ref{pt:NoLiveGoroutines}              & Yes     & Yes          & Yes          & Yes          \\
\ref{pt:NoSender}                      & Yes     & Yes          & Yes          & Yes          \\
\ref{pt:out-of-order}                  & Yes     & Yes          & Yes          & Yes          \\
\ref{pt:return-channel}                & Yes     & Yes          & \textbf{Err} & Yes			\\\hline
cockroachdb\#25456 & Yes     & Yes          & \textbf{Err} & Yes          \\
grpc-go\#1424      & Yes     & Yes          & Yes          & Yes          \\
moby\#33293        & Yes     & \textbf{Err} & \textbf{Err} & Yes          \\
moby\#4395         & Yes     & Yes          & \textbf{Err} & Yes
\end{tabular}
\end{table}

Although the Flow extension works well for the selected concurrency features,
GoDDaR, GOMELA, and GoAT can very well excel at features that we did not test this time,
including buffered channels, channel closing, the select statement, and loops.

\section{Limitations}\label{sec:limitation}

From the syntactic aspect,
we only support in this paper unbuffered channels, bidirectional channels, and the keyword \code{go} and \code{defer}.
We recognize a limited number of built-in APIs that return a channel, such as \code{time.After}.
When other features are present in a program, such as receive-only or send-only channels,
the Flow type system raises an error.
In this case, other deadlock analyzers are needed for this program.

An important limitation is from the semantic aspect.
Coroutine Types as well as our Flow extension model the collective behavior of a set of coroutines by using types.
Because a type is expressed as $\tau / \rho$,
our Flow type system maps the type of a concurrency expression onto $\tau$,
and maps the guarded if-conditional expression onto $\rho$, bridging the pure type-based analysis and pure value-based analysis.

\begin{lstlisting}[language=go, float=ht, label=out-of-order2, caption={We currently do not distinguish two channels with the same type}]
func work(cInt chan int, cStr chan int) {
	fmt.Println(<-cInt)
	fmt.Println(<-cStr)
}

func main() {
	cInt := make(chan int)
	cStr := make(chan int)
	go work(cInt, cStr)

	cStr <- 2
	cInt <- 1
}
\end{lstlisting}

Nevertheless, we do not track channel names and variables outside of if-conditionals in this paper.
To illustrate, Listing~\ref{out-of-order2} is a modification of Listing~\ref{out-of-order} and involves two channels of int.
The nature of the problem does not change; it still runs into a deadlock.
Our Flow type system assigns $\corD{{\Yield{\tlit{Int}}}{\Yield{\tlit{Int}}}}$ to \code{work()}
and assigns  $\corD{{\Start(work)}{\Receive{\tlit{Int}}}{\Receive{\tlit{Int}}}}$ to \code{main()}.
As channel names are stripped, the reduction algorithm sees the two yielding and two receiving operations are properly paired,
and (wrongly) considers the program as free of deadlocks.
One fix is probably to add a label denoting the channel object into the Flow items.
We believe the underlying coroutine reduction algorithm is capable of handling that,
so we leave it as future work to get the ball rolling.

Finally, from the operational aspect,
because we track variable names in the $\rho$ part for evaluating conditionals, we could suffer from the state-explosion problem as other static analysis tools.
As we have seen in Listing~\ref{conditional}, a single variable \code{weekday} expands to 4 states.
With many variables, a large program can be hard to reason about.
In the future, we plan to use an SMT solver (such as Z3) to reason the whole program, rather than only the $\rho$ part,
because an SMT solver is good at handling thousands or millions of variables~\cite{abraham2017smt}.
After all, a deadlock may only depend on a couple of key variables.




\section{Conclusion}\label{sec:conclusion}

A coroutine is a programming construct that makes code easier to understand and manage, and can run in parallel.
Coroutine Types~\cite{gu2023typing} are behavioral types to model interactions of coroutines with a single receiving operation followed by a single yielding operation.
We contribute a Flow extension to Coroutine Types, so that coroutines with more than one receiving and yielding operation can be modeled.
We accordingly revise the reduction rules of Coroutine Types.
To show the usefulness of the Flow extension, we contribute a Flow type system that maps {\Golang} expressions to Coroutine Types,
while also considering values in \code{if} statements.
The Flow type system is our attempt to bridge the pure type-based analysis and pure value-based analysis.
After we reduce Coroutine Types, if any yielding and receiving part remains, the program will manifest blocking behaviors.

Go is a popular programming language for distributed systems,
but a frequent kind of bugs is deadlocks due to the wrong use of concurrency features.
In this paper, we concentrate on the most commonly used semantics in Go: unbuffered channels with the \code{go} and \code{defer} keywords.
If a {\Golang} file uses unsupported features, our tool gives a warning and refuses to analyze.

We collected {\TestsCount} pieces of both positive and negative code for evaluation purposes
because a subtle edit could fix a blocking behavior.
Our algorithm was designed to identify 17 kinds of interactions, and furthermore successfully predicted deadlock behaviors in 4 real-world programs.
In contrast, three state-of-the-art deadlock analyzers crashed or gave wrong predictions in some cases.
Therefore, our type-based deadlock analyzer not only fills the gap in the landscape of value-based detection, but also complements existing detectors.

In the future, we would like to improve Coroutine Types with Flow to express more language constructs, such as buffered channels in Go or other programming languages.
We are also eager to formally study the liveness and safety properties of processes.
Finally, rather than using Microsoft Z3 for the sole purpose of evaluating $\rho$,
we plan to use Z3 more extensively, such as for finding condition combinations where a concurrency bug occurs.

\section*{Data Availability}
We published our source code at \cite{source-code-anonymous}.
The Flow extension is in the folder \code{GeneratorCalculation};
the Flow type system in the folder \code{Go};
the pattern files and the test cases are in the folder \code{GoTests};
the way to build and run the program in the folder \code{.github/workflows}.

Please consult \cite{gu2024typing} about the source code of the base Coroutine Types.

\bibliographystyle{IEEEtran}
\bibliography{ref}

\end{document}